\documentclass[lettersize,journal]{IEEEtran}
\usepackage{amsmath,amssymb,amsfonts}
\usepackage{algorithmic}
\usepackage{graphicx}
\usepackage{subfigure}
\usepackage{textcomp}
\usepackage{xcolor}
\usepackage{graphicx}
\usepackage{epstopdf}
\usepackage{multirow}
\usepackage{array}
\usepackage{cite}
\usepackage{amsmath}
\usepackage{makecell}
\usepackage{tabularx}
\usepackage{pdfpages}
\usepackage[margin=10pt,font=small]{caption}
\usepackage[ruled]{algorithm2e}
\def\BibTeX{{\rm B\kern-.05em{\sc i\kern-.025em b}\kern-.08em
    T\kern-.1667em\lower.7ex\hbox{E}\kern-.125emX}}

\begin{document}


\title{Channel Sparsity Variation and Model-Based Analysis on 6, 26, and 132 GHz Measurements}

\author{Ximan Liu, Jianhua Zhang, Pan Tang, Lei Tian, Harsh Tataria, Shu Sun, and Mansoor Shafi\vspace{-20pt}}



\maketitle

\begin{abstract}
Triggered by the introduction of higher frequencies (above 24 GHz), there has been a long-standing debate in the radio propagation community on whether higher frequency radio channels are sparser relative to channels below 6 GHz. Here, sparsity implies a few dominant multipath components containing the vast majority of the electromagnetic energy. This discussion has recently been revisited with the study and interest in bands above 100 GHz for future wireless access. In this paper, the level of sparsity is examined at 6, 26, and 132 GHz carrier frequencies by conducting channel measurements in an indoor office environment. By using the Gini index (value between 0 and 1) as a metric for characterizing sparsity, we show that increasing carrier frequency leads to increased levels of sparsity. The measured channel impulse responses are used to derive a Third-Generation Partnership Project (3GPP)-style propagation model, used to calculate the Gini index for the comparison of the channel sparsity between the measurement and simulation based on the 3GPP model. Our results show that the mean value of the Gini index in measurement is over twice the value in simulation, implying that the 3GPP channel model does not capture the effects of sparsity in the delay domain as frequency increases. In addition, a new intra-cluster power allocation model based on measurements is proposed to characterize the effects of sparsity in the delay domain of the 3GPP channel model. The accuracy of the proposed model is analyzed using theoretical derivations and simulations. Using the derived intra-cluster power allocation model, the mean value of the Gini index is 0.97, while the spread of variability is restricted to 0.01, demonstrating that the proposed model is suitable for 3GPP-type channels. To our best knowledge, this paper is the first to perform measurements and analysis at three different frequencies for the evaluation of channel sparsity in the same environment.
\end{abstract}

\begin{IEEEkeywords}
3GPP channel model, Gini index, LoS, propagation measurements, sparsity.  
\end{IEEEkeywords}

\section{Introduction}
\IEEEPARstart 
{T}{he} future mobile communications will require higher data rates and more abundant spectrum resources \cite{b1}. To meet the needs of future mobile communications, the sub-terahertz (sub-THz) band is considered as an important range of frequencies to support some of the envisioned use cases. These use cases and their performance requirements are discussed in \cite{b2}, and many references are cited therein.

The THz band generally refers to the frequency range of 1 to 10 THz and the sub-THz band is a smaller subset of this range, i.e., 0.1 to 1 THz. The benefit of this band is the relative availability of abundant spectrum resources needed for the use cases that need terabits/second data rates. However, there are also many challenges to the use of this band consisting of very large signal attenuation, reduced degrees-of-freedom, limits of devices and semiconductors, to name a few. A comprehensive list of the challenges in using these very high-frequency ranges and some possible solutions is discussed in \cite{b2}. All radio channels (regardless of the frequency band of operation) are considered as sparse \cite{b3,b4,b23}, as practical systems have limited bandwidth. Nonetheless, it is widely believed that millimeter-wave (mmWave) channels are more sparse relative to centimeter-wave (cmWave) systems \cite{b5}. Therefore, it is reasonable to conjecture that the sparsity of sub-THz bands will be even higher than the mmWave bands. In the case of mmWave systems the signal is more likely to be obscured due to the short wavelength \cite{b6,b7,b30}. We expect similar behavior in the sub-THz range. The large diffraction losses arise due to the narrowing of the Fresnel zone at increasing frequencies. This in turn means that only the line-of-sight (LoS) path and specular reflections are the main mechanisms of propagation between the transmitter and the receiver \cite{b6,b7,b8,b31}. There may be cases when the LoS path is blocked leaving only the reflected paths (some of which may be dominant) as the source of propagation.

Therefore, the channel will be more likely to exhibit sparsity. Whilst this is true about the mmWave channels \cite{b7,b8}, can the same be concluded about the sub-THz spectrum? A correct estimation of sparsity is important in the design of cellular systems \cite{b5,b9,b25,b27}. Especially in vehicle-to-vehicle scenarios, the beamforming algorithms rely on the channel sparsity. What then is a measure of sparsity? There are many measures of sparsity - such as \cite{b10}: 
 \begin{itemize}
 \item Gini index that arises from measuring economic disparities, 
 \item the Ricean K factor - a measure of the concentration of power in the LoS ray,
 \item spatial degrees of freedom.
 \end{itemize}
 
 The most common definition of sparsity is interpreted as when a signal consists of $s$ dominant signal vectors and most of the power is concentrated in the dominant signal vectors \cite{b24}. But there is also a view that in practical mmWave systems, the narrow beamwidth of the antenna arrays used contributes to sparsity and just a few paths can be detected. A universal definition of sparsity is not there. 
 
 A recent paper \cite{b11} has shown via measurements that mmWave channels are sparser relative to cmWave channels. But measurements to describe the sparsity of sub-THz channels and compare this with lower frequencies are still lacking.

In recent years, some channel measurements have been designed to study channel sparsity. In \cite{b12}, the measurement-based channel characterization is presented in a large hall scenario at 299-301 GHz. The results in the LoS scenario illustrate the sparsity of the multipath. In \cite{b13}, a measurement campaign was conducted at 145 GHz in an outdoor urban environment. It was found that it is not entirely reasonable to measure sparsity from only a small number of multipath components (MPCs), since the measurements show that there are abundant MPCs. In \cite{b14}, a measurement campaign was conducted in a typical vehicular urban scenario to analyze the channel sparsity by many measures. The authors evaluate some common measures to describe sparsity and conclude that the Gini index \cite{b15} and degrees of freedom (DoF) \cite{b16} are good choices because they predict sparsity more accurately. In \cite{b17}, the authors compare and analyze the channel sparsity in the corridor, workroom, and factory to study the effect of the environment. The channels in the corridor have the weakest sparsity. This is because the narrow walls provide a large number of reflected rays. The factory has sparser channels because of the complex environment. However, the measurement experiments of channel sparsity in sub-THz bands are not yet abundant. There are few measurement experiments to verify that THz channels are sparser than mmWave channels. The indoor office scenarios as the most likely deployment of the sub-THz bands use cases (i.e., holograms). Therefore, in this paper we performed the measurement in an indoor scenario for three frequency bands, in the cmWave, mmWave, and sub-THz range, and compare their sparsity.

The Third-Generation Partnership Project (3GPP) channel model is widely used in the research, and development of communication systems \cite{b18}. This model is limited to frequencies below 100 GHz and even for mmWave bands is not able to characterize sparsity. In fact, the sparsity of the channel matrix derived from the 3GPP channel model is similar regardless of the frequency band of operation - a rather serious limitation of this model. Furthermore, the only occasion a 3GPP model derived channel displays sparsity properties is when there is a LoS path. Therefore, the 3GPP channel model for the non-line-of-sight (NLoS) paths has deficiency in characterizing sparsity, which leads to inaccurate prediction results. We observe this by comparing the Gini index from the measurement and that computed from a 3GPP model fitted to the measurements. The deficiency of the 3GPP model to characterize sparsity arises from the model allocating cluster power equally to the rays within the cluster. Clearly, sparsity is a key feature in channel models and this drawback needs to be addressed.

The purpose of this paper is to analyze and model the sparsity in cmWave, mmWave, and sub-THz bands by multi-band measurements. The sparsity study provides new insights for beam study in vehicular communication systems. The channel measurements in cmWave, mmWave, and sub-THz bands were performed in the same indoor scenario to verify the sparsity characteristics. The computation of the cumulative density function (CDF) of the Gini index (our chosen metric of sparsity) from the measurements is also given to more clearly represent the sparsity variation in different frequency bands. The deficiencies of the 3GPP model are addressed by proposing a new intra-cluster power allocation model that recovers the sparsity.
The main contributions of this paper are as follows:
\begin{itemize}
\item From the multi-band channel measurements in the same indoor office scenario performed in cmWave, mmWave, and sub-THz bands, we conclude that the Gini index shows that the channel sparsity in sub-THz bands is more significant than that of in the cmWave and mmWave. These results show that the sparsity level increases with frequency.
\item The sparsity of simulated and measured channels is compared to demonstrate that the 3GPP channel model lacks the ability to characterize sparsity. The Gini index simulated by the 3GPP channel model does not show the same sparsity level. The reconstructed Gini index is generated by simulation experiments based on the 3GPP channel model. The large-scale parameters of the 3GPP channel model are extracted from the measurement results in sub-THz bands.
\item We present a method to address the deficiencies of the 3GPP channel model to characterize sparsity in the delay domain. Based on the measurement results, a new intra-cluster power allocation model is proposed. The modified 3GPP channel model with the intra-cluster power allocation model has the ability to characterize channel sparsity in the delay domain. The performance of the intra-cluster power allocation model is analyzed by theoretical derivation and simulation experiments. The simulation results demonstrate that the Gini index generated by the modified model with intra-cluster power allocation model is closer to the real channel.
\end{itemize}

This paper is organized as follows: Section \ref{II} describes the setup of the measurement and simulation experiments in cmWave, mmWave, and sub-THz bands. The data processing methods are also presented. Section \ref{III} analyzes the sparsity using the measurement results. In addition, it is shown that the 3GPP channel model lacks the ability to characterize sparsity in the delay domain by comparing the measurement and simulation results. Therefore, a sparsity model is proposed based on the 3GPP channel model in Section \ref{IV}. A new intra-cluster power allocation model is proposed to enable the 3GPP channel model to characterize sparsity. The reliability of the sparsity model is verified in Section \ref{V} using theoretical derivation and simulation results. The conclusion of this paper is provided in Section \ref{VI}.

\section{Measurement and Simulation}\label{II}
In this section, the measurement system setup and parameters at three different frequencies are given. Furthermore, the methods for extracting MPCs for the indoor office scenario are presented. Based on the measurement, the Gini index from the 3GPP model is derived in order to compare the modelled results with the measurements. The large-scale parameters i.e., the number of clusters, delay spread, and Ricean K-factor are obtained by clustering results, as explained later in the section. 

\subsection{Measurement Setup and Data Processing}

\begin{figure}[tbp]
\centering
\subfigure[]{\includegraphics[scale=0.4]{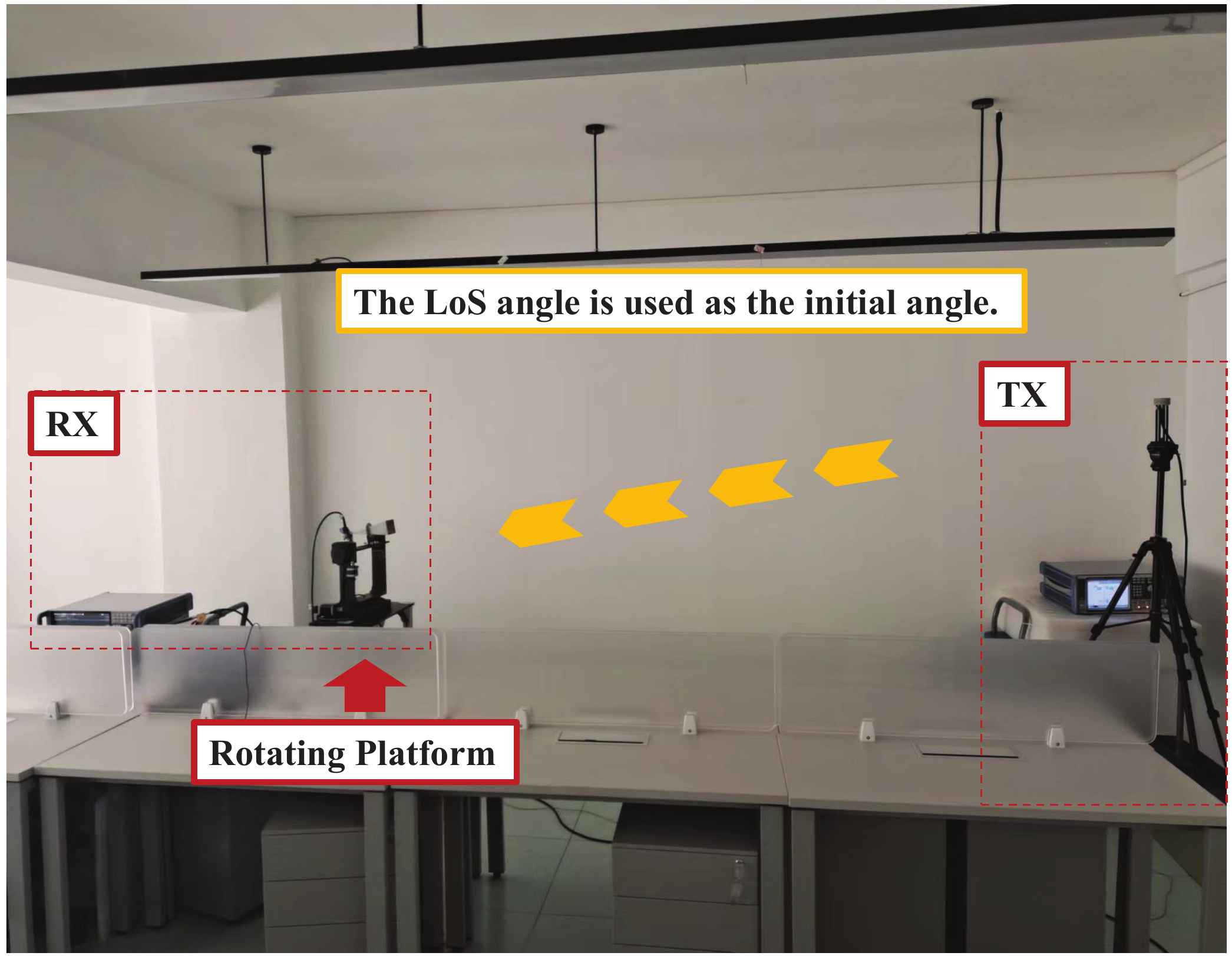}
\label{TX}}
\subfigure[]{\includegraphics[scale=0.4]{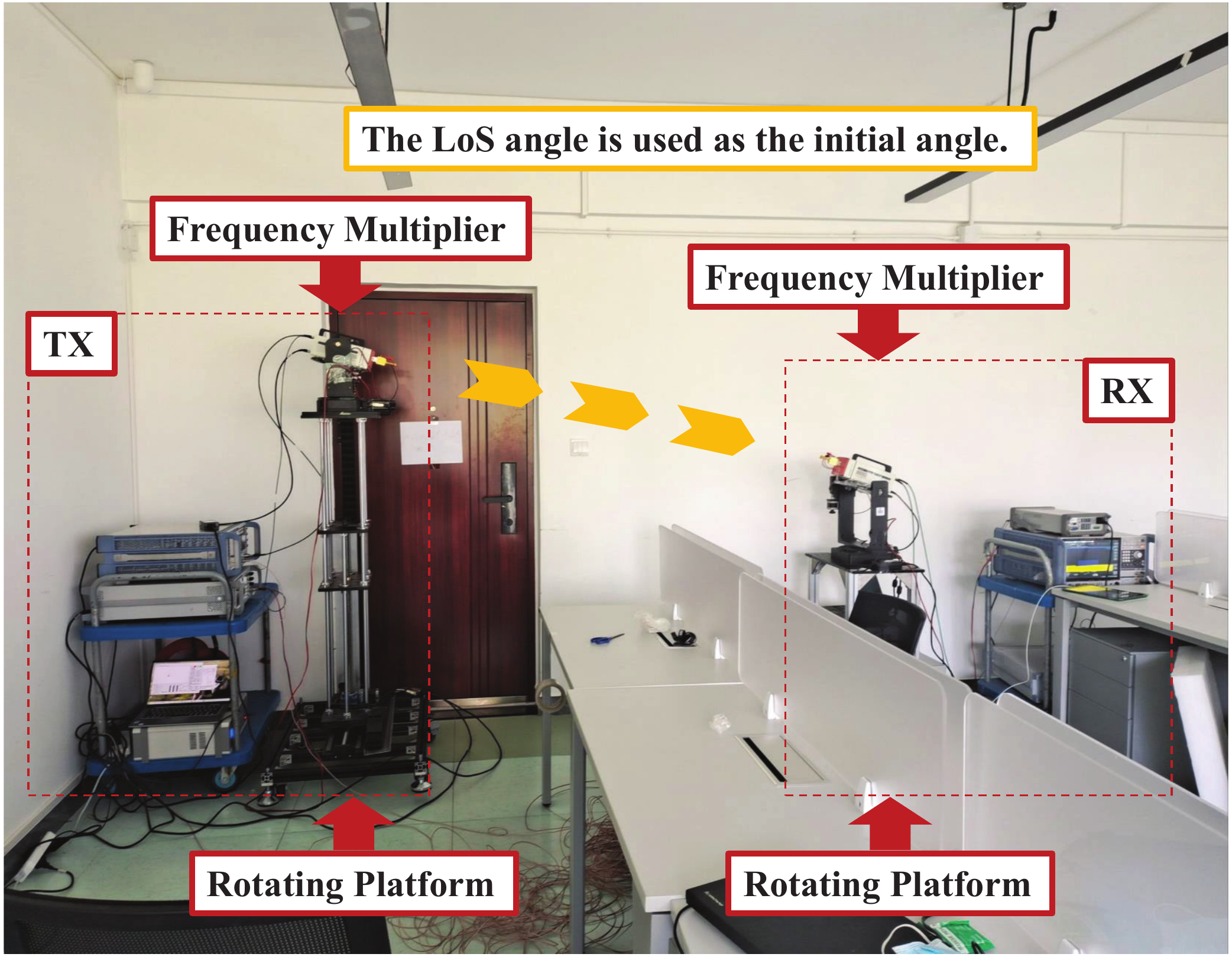}
\label{TXTHz}}
\caption{The measurement system. (a) In cmWave and mmWave bands. (b) In sub-THz bands.}
\label{ms}
\end{figure}

The channel measurements were performed via a time domain channel sounding platform in the indoor office scenario in cmWave, mmWave, and sub-THz bands. In the cmWave and mmWave channel measurements, an omnidirectional antenna is used for the transmitter (TX) and a horn antenna is used for the receiver (RX), as illustrated in Fig.~\ref{TX}. The TX system is composed of a signal generator and an omnidirectional antenna. The signal generator outputs a signal to the omnidirectional antenna to transmit the signal. The RX system is composed of a directional antenna, a rotary table, a low-noise amplifier and a spectrum analyzer. The directional antenna is on a platform that can be rotated in elevation and azimuth. The directional antenna receives the signal to the low-noise amplifier and then inputs it into the spectrum analyzer for signal analysis. The discrete-time channel impulse responses (CIRs) are obtained by sampling the continuous-time received signal. In the sub-THz channel measurement, it is necessary to add a frequency multiplier and a local oscillation source to the original measurement system due to the high-frequency band. Because of the significant attenuation in sub-THz bands, the horn antennas are used for both TX and RX to enhance the signal, as illustrated in Fig.~\ref{TXTHz}. The channel status information with complete angles is obtained by rotating the horn antenna.

\begin{figure}[tbp]
\centerline{\includegraphics[scale=0.55]{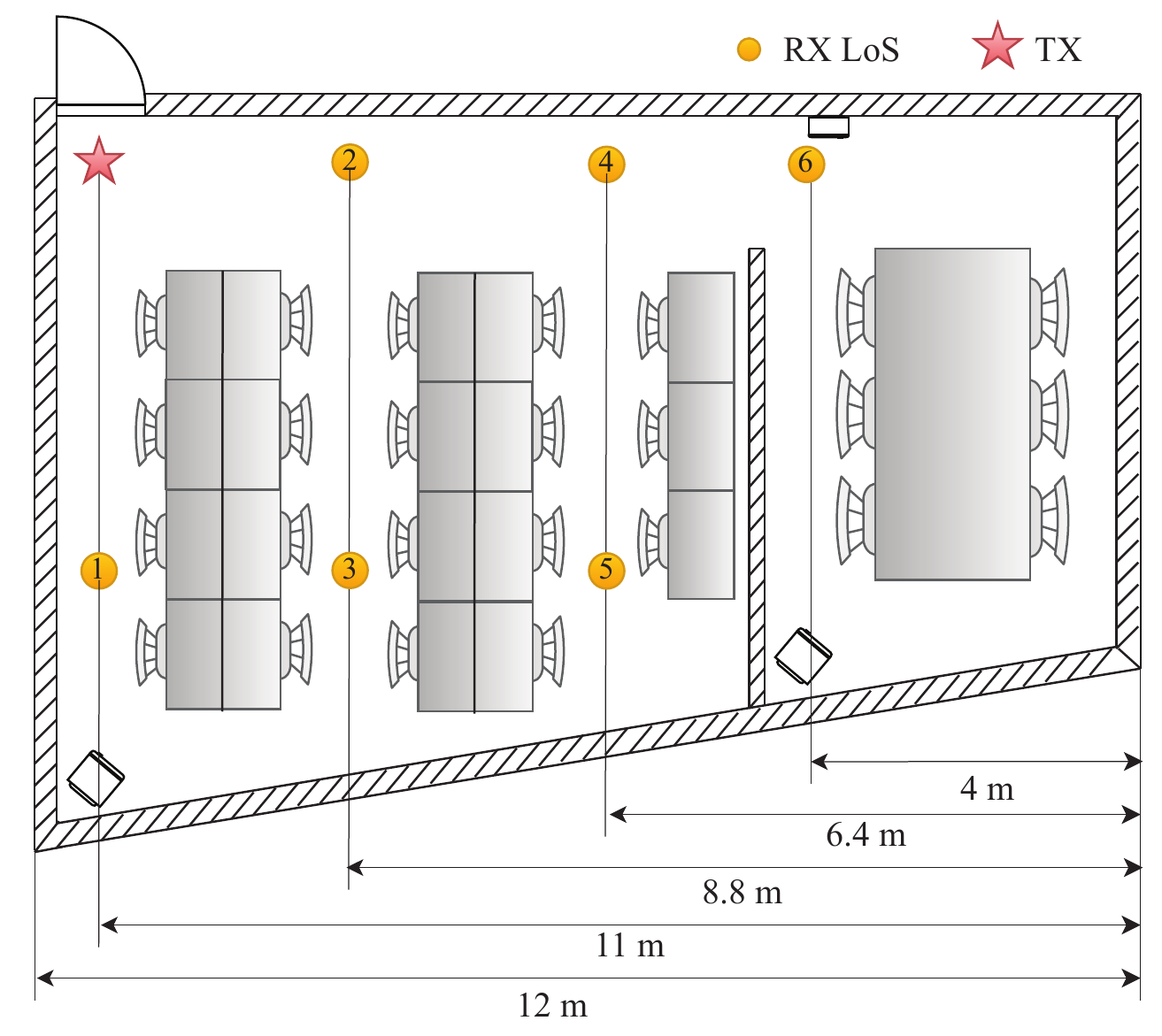}}
\caption{Measurement Scenario.}
\label{scenario}
\end{figure}

The measurement scenario is illustrated in Fig.~\ref{scenario}. The measurement scenario is a typical indoor office scenario with an area of 84 square meters. The red star represents the location of TX and the yellow circulars represent the location of RX in LoS cases. A wall with a length of 5 m is in the middle of the office. These positions have complete angular information by the antenna rotation. The parameters of measurement system are illustrated in Table~\ref{parameters}.

\begin{table}[htbp]
\caption{Measurement Parameters}
\begin{center}
\begin{tabular}{|p{1.9cm}<{\centering}|p{1.7cm}<{\centering}|p{1.7cm}<{\centering}|p{1.8cm}<{\centering}|}
\hline
\textbf{Parameters}&\multicolumn{3}{c|}{\textbf{Value/Type}} \\
\cline{2-4} 
\hline
Scenario&\multicolumn{3}{c|}{Indoor Office} \\
\hline
\makecell[c]{Carrier\\Frequency}& 6 GHz & 26 GHz & 132 GHz\\
\hline
\makecell[c]{RF\\Bandwidth}&\multicolumn{2}{c|}{200 MHz} & 1.2 GHz\\
\hline
Antenna Type &\multicolumn{2}{c|}{\makecell[c]{TX: Omnidirectional antenna,\\RX: Horn antenna}} & \makecell[c]{TX and RX:\\Horn antenna}\\
\hline
Antenna Height&\multicolumn{3}{c|}{TX: 1.8 m, RX: 1.2 m} \\
\hline
\multirow{2}{*}{\makecell[c]{3 dB Bandwidth\\in Azimuth}}
& \multirow{2}{*}{15.5 deg}  &\multirow{2}{*}{9.02 deg}  & TX: 14 deg\\
\cline{4-4}
&                            &                           &RX: 9.9 deg\\
\hline
\multirow{2}{*}{Antenna Gain}
& TX: 4 dBi    &TX: 4 dBi  & TX: 23.25 dBi\\
\cline{2-4}
& RX: 20.3 dBi & RX: 24.75 dBi &RX: 25.1 dBi\\
\hline
TX Signal &\multicolumn{3}{c|}{Pseudorandom Noise Sequence of Order 9} \\
\hline
Transmit Power &\multicolumn{3}{c|}{0 dBm} \\
\hline
\end{tabular}
\label{parameters}
\end{center}
\end{table}

According to the hotspot carrier frequencies for cmWave, mmWave, and sub-THz, this paper use 6, 26, and 132 GHz to study the sparsity. The height of the RX antenna is set higher than the desktop height to mimic the wireless communication system of the real office scenario. All of the measurements were performed in the static scenario. The horn antenna was rotated with the LoS angle as the origin to obtain the complete angle information at the RX side. In the cmWave and mmWave channel measurements, the TX antenna is fixed and the azimuth angles of RX antenna rotation are in steps of $10^\circ$ from $-180^\circ$ to $180^\circ$. The elevation angles are  $-10^\circ$, $0^\circ$, and $10^\circ$. To obtain complete angle information, the step size is related to the antenna parameters, as illustrated in Table~\ref{parameters}. In the sub-THz channel measurement, the antenna at TX also needs to be rotated to obtain channel information with complete angles since the TX antenna is a directional antenna. Note that, this paper is interested in the likelihood of a dominant ray capturing most of the energy at different frequencies. Therefore, the analytical results of sparsity are convincing when the temporal resolution of the high carrier frequency is not lower than that of the low carrier frequency.


The effective rays are estimated from the raw CIRs with complete angles \cite{b19}. The power delay profile (PDP) is calculated from the CIR. The calculation of the peaks for each PDP is that we iterate through all the delay bins and determine the peaks when the power is greater than both the previous and next one. For two adjacent closely-spaced peaks, their time interval needs to be larger than the temporal resolution of the measurement system so as to be regarded as two independent rays. Each PDP corresponds to an antenna angle. Therefore, the time delay, power, and angle information of each ray can be calculated from CIRs. The lobe pattern of the antenna may cause the RX to receive the same ray at a similar antenna angle. Because of the antenna response, these same rays have the same time delay and different power. The MPCs with the same time delay are first found in the CIRs of similar angles. Afterward, a complete CIR is synthesized by screening the MPCs with the strongest power among the MPCs with the same delay. Each position can obtain 100 CIRs with complete angle information. The peaks in each PDP after screening are extracted as effective rays, as elaborated above. The power of the effective rays will be used for the study of sparsity.

\subsection{Simulation of Gini index Derivation Based on 3GPP model}\label{IIB}
The simulation experiments derive the Gini index based on the 3GPP channel model by the following steps \cite{b18}. Whilst the 3GPP channel model is limited to frequencies below 100 GHz, we have assumed the same method to fit model parameters to obtain channels in sub-THz bands.

The first step sets the parameters of the scenario and antennas based on measurement. The second step generates some large-scale parameters, e.g. delay spread (DS), Ricean K-factor, and the number of clusters. To compare the simulation results with the measurement fairly, the parameters of the 3GPP channel model are obtained by measured clustering results. The clustering algorithm considers the delay, angle, and power. The parameters of the 3GPP model are calculated by the method mentioned in \cite{b20}, as illustrated in Table~\ref{Largescale}. The third step generates the delay for each cluster based on the DS shown in Table~\ref{Largescale}. The fourth step uses the delay to generate the power of each cluster. The power of rays within a cluster is all the same. The power of all rays is superimposed to obtain the power of the cluster. 

\begin{table}[htbp]
\caption{The Large-Scale Parameters.}
\begin{center}
\begin{tabular}{|p{2cm}<{\centering}|p{1.6cm}<{\centering}|p{1cm}<{\centering}|p{1cm}<{\centering}|p{1cm}<{\centering}|}
\hline
\multicolumn{2}{|c|}{Parameters}  &     cmWave   &    mmWave   &    sub-THz\\
\hline
 \multirow{2}{*}{$\log_{10}{\left ( DS/1s \right ) } $}
             &       $\mu $       &     -7.17    &     -7.42   &    -8.47\\
\cline{2-5} 
             &      $\sigma $     &      0.4     &      0.46   &     0.67\\
\hline
 \multirow{2}{*}{\makecell[c]{Ricean K-factor\\(dB)}}
             &       $\mu $       &      4.23    &     5.52    &     8.0\\
\cline{2-5} 
             &      $\sigma $     &      3.25    &     4.36    &     7.9\\
\hline
Cluster Number&      $\mu $       &        9     &      8      &     3\\
\hline
\end{tabular}
\label{Largescale}
\end{center}
\end{table}

 At last, the Gini index is reconstructed according to the power and the number of rays obtained by the 3GPP channel model \cite{b18,b14}. 

\section{Channel Sparsity Characterization}\label{III}



This section analyzes the sparsity by comparing the Gini index of measured results in cmWave, mmWave, and sub-THz bands. The Gini index has been considered in economics in the study of inequities in the distribution of wealth \cite{b15}. It meets the definition of sparsity, where a few elements contain a large proportion of the total power. The Gini index as a sparsity measure satisfies Dalton's 1st Law \cite{b26}. The Gini index has been widely used in recent years in the study of channel sparsity \cite{b21,b22,b28}. The Gini index ranges from $\left [ 0,1 \right ] $ with no units. The closer the Gini index is to 1, the sparser the channel is. When the Gini index is equal to 0, the power of all rays in the channels is equal. This paper analyzes the channel sparsity at cmWave, mmWave, and sub-THz using the Gini index, expressed as \cite{b14}

\begin{equation}
G=1-2\sum_{i=1}^{R}\frac{p_i}{\left \| \mathbf{p}   \right \|_1 }  \left ( \frac{R-i+\frac{1}{2} }{R}  \right ), 
\label{Gini}
\end{equation}

\noindent where $R$ represents the number of rays, $p_i$ represents the power of the $i$-th ray. $\mathbf{p}$ represents a power vector composed of the power of $R$ rays. The elements in $\mathbf{p}$ are arranged in an ascending order, i.e., $p_{1} < p_{2} < \cdots  < p_{R} $. The subscripts are after sorting.

To study the sparsity in cmWave, mmWave, and sub-THz bands, the channel measurements were performed at six positions in this paper. To investigate the characteristics of channel sparsity with frequency, the channel measurements were performed at the same positions in the three frequency bands. The measurement process remains stationary. The Gini index is calculated using the impulse responses of a large number of samples. 

\begin{figure*}[!htbp]
\centering
\subfigure[]{
	\label{fig:subfig:a} 
	\includegraphics[width=4.5cm,height=3.5cm]{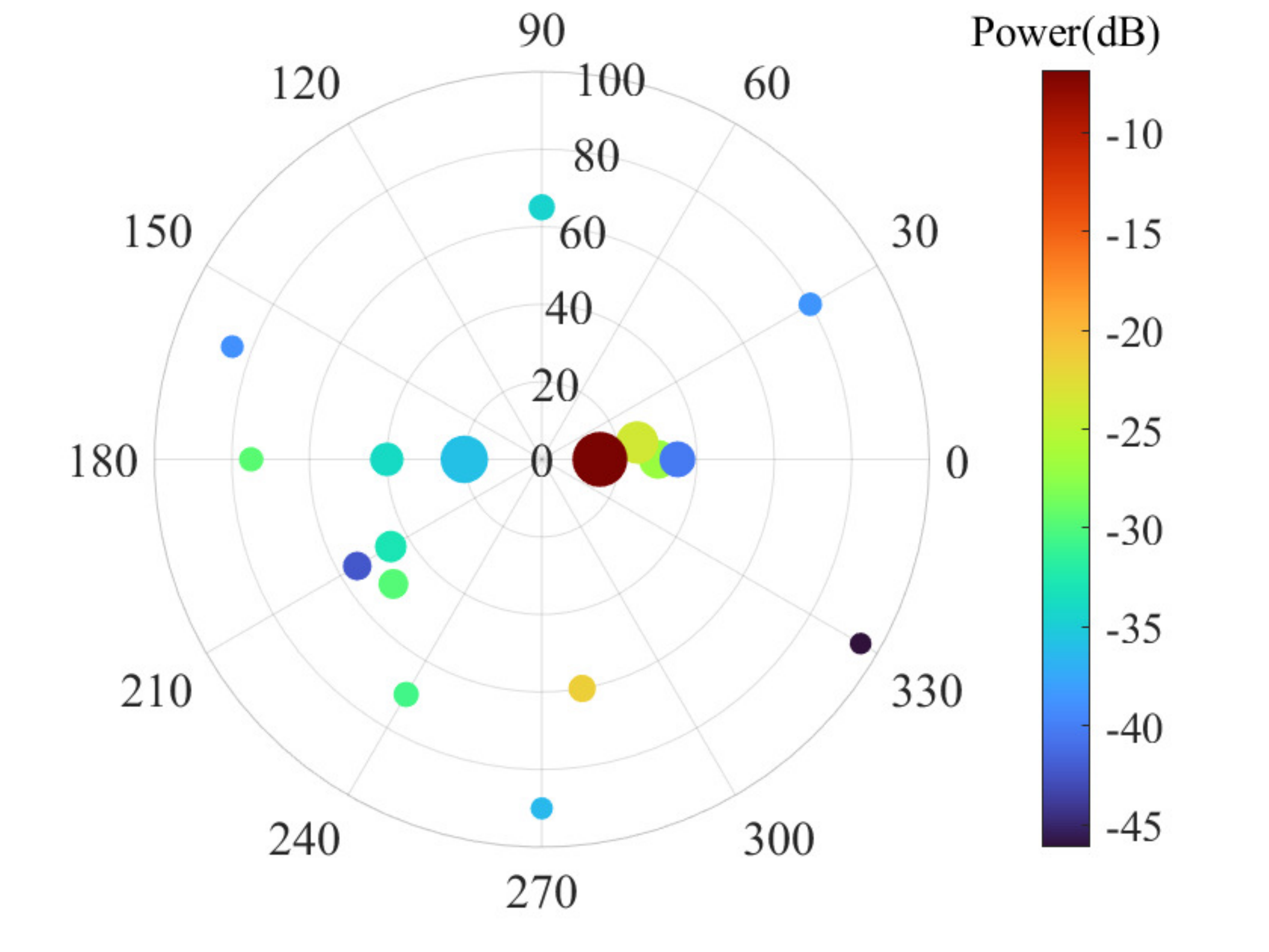}}
\hspace{0.5in}
\subfigure[]{
	\label{fig:subfig:b} 
	\includegraphics[width=4.5cm,height=3.5cm]{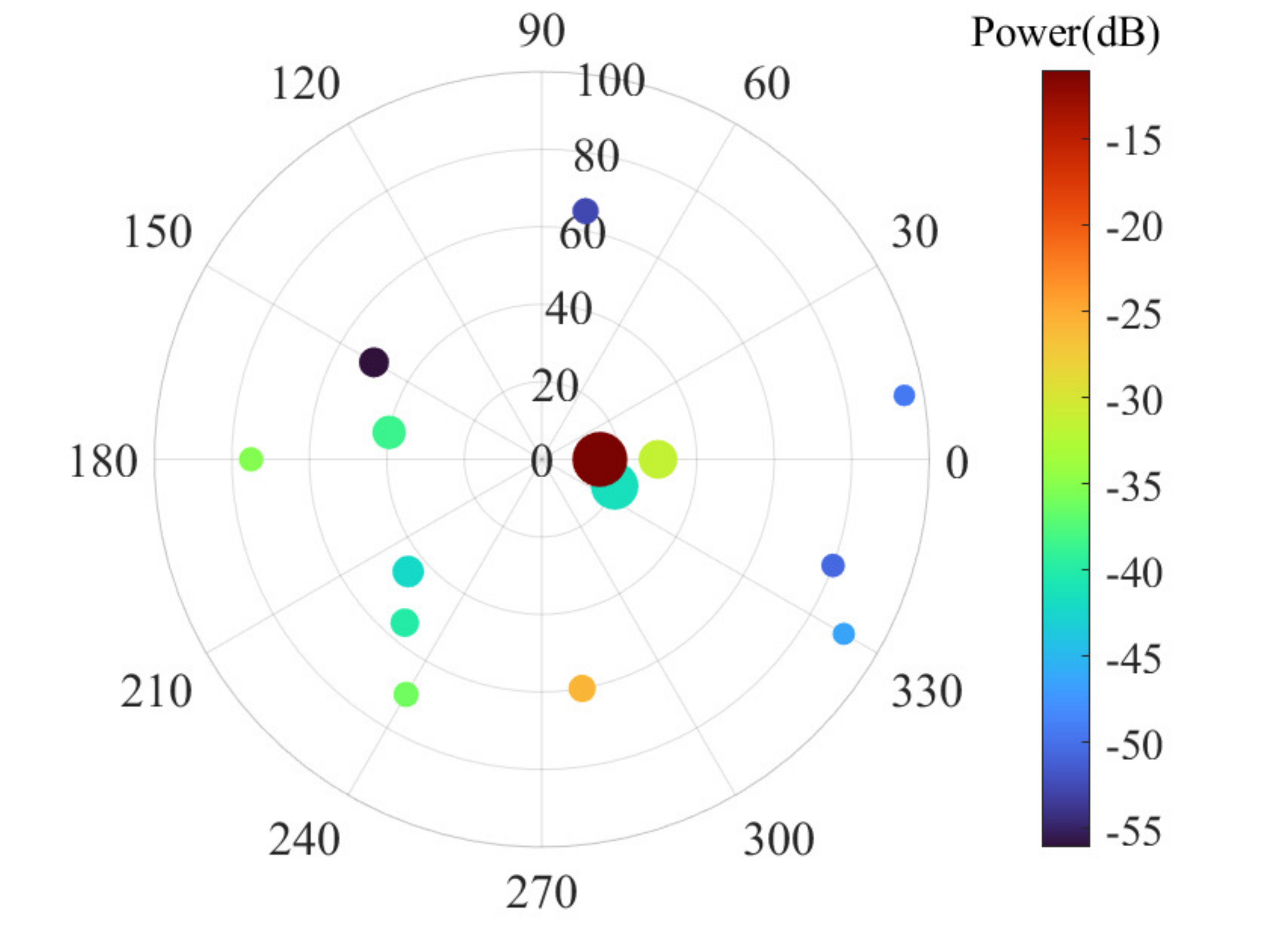}}
\hspace{0.5in}
\subfigure[]{
	\label{fig:subfig:c} 
	\includegraphics[width=4.5cm,height=3.5cm]{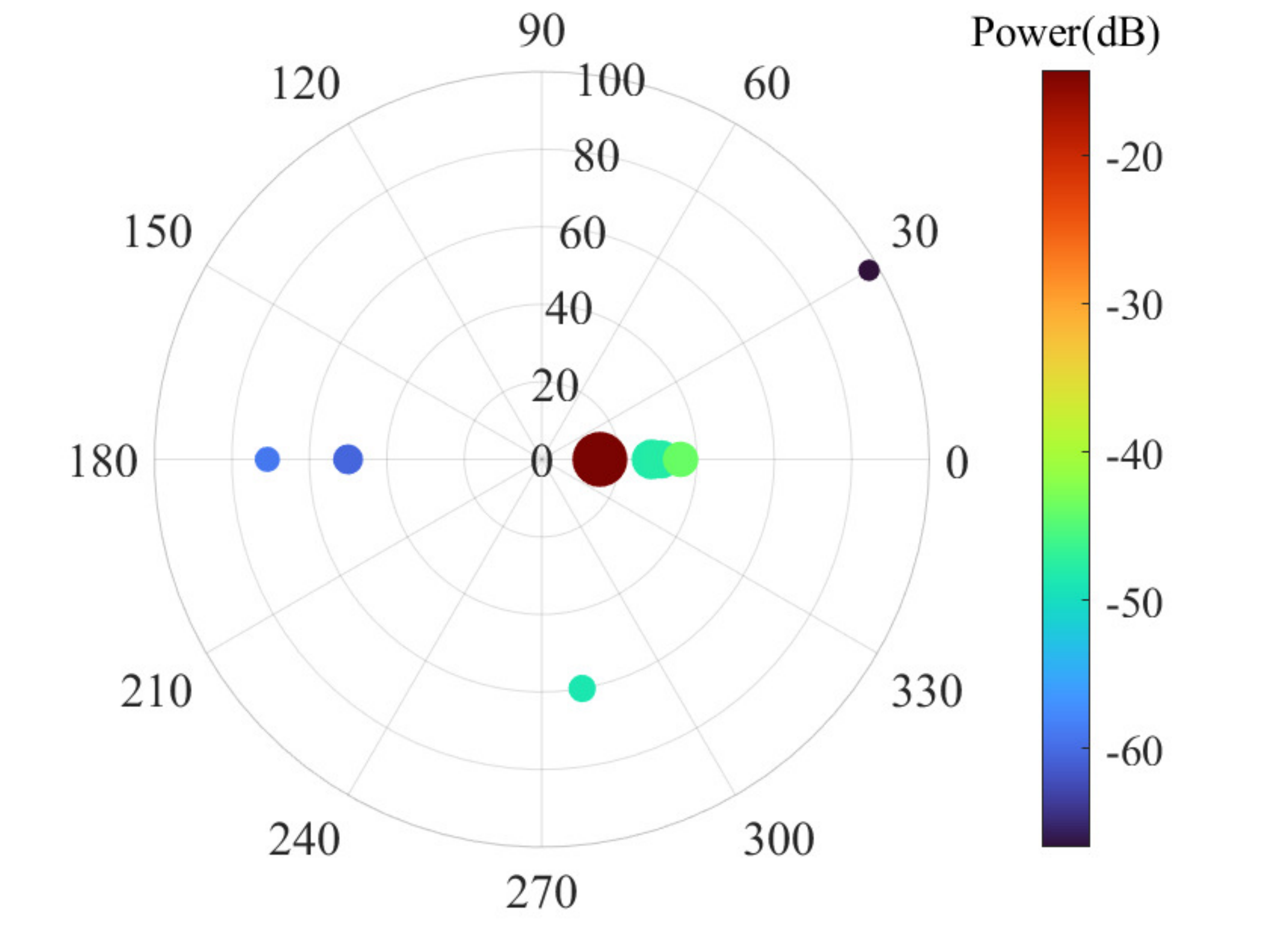}}
\caption{The delay, power, and AOAs of the rays in the 2nd position at cmWave, mmWave, and sub-THz. The unit of delay is nanoseconds. (a) In cmWave bands. (b) In mmWave bands. (c) In sub-THz bands.}
\label{PDP} 
\end{figure*}

The power angle spectrum (PAS) of effective rays is analyzed by taking snapshots of the complete channels at each position. From the PAS of the 2nd position in cmWave, mmWave, and sub-THz bands illustrated in Fig.~\ref{PDP}, it is obvious that the number of rays increases significantly from sub-THz to cmWave bands. The RX received some rays that traveled long distances in low-frequency bands. These rays may have multiple reflection or diffraction. However, the sub-THz channels are not sensitive to these rays because of significant loss. The THz waves have high reflection loss, affecting the NLoS components \cite{b29}. The power of the LoS path gradually decreases from cmWave to sub-THz bands due to the propagation loss characteristics of the THz channels. The proportion of power contained in the dominant ray also decreases gradually with decreasing frequency. This is because the significant reflection loss of THz communication leads to the weak power of NLoS rays. It is proved that the channels in sub-THz bands are sparser.

\begin{figure}[tbp]
\centerline{\includegraphics[scale=0.3]{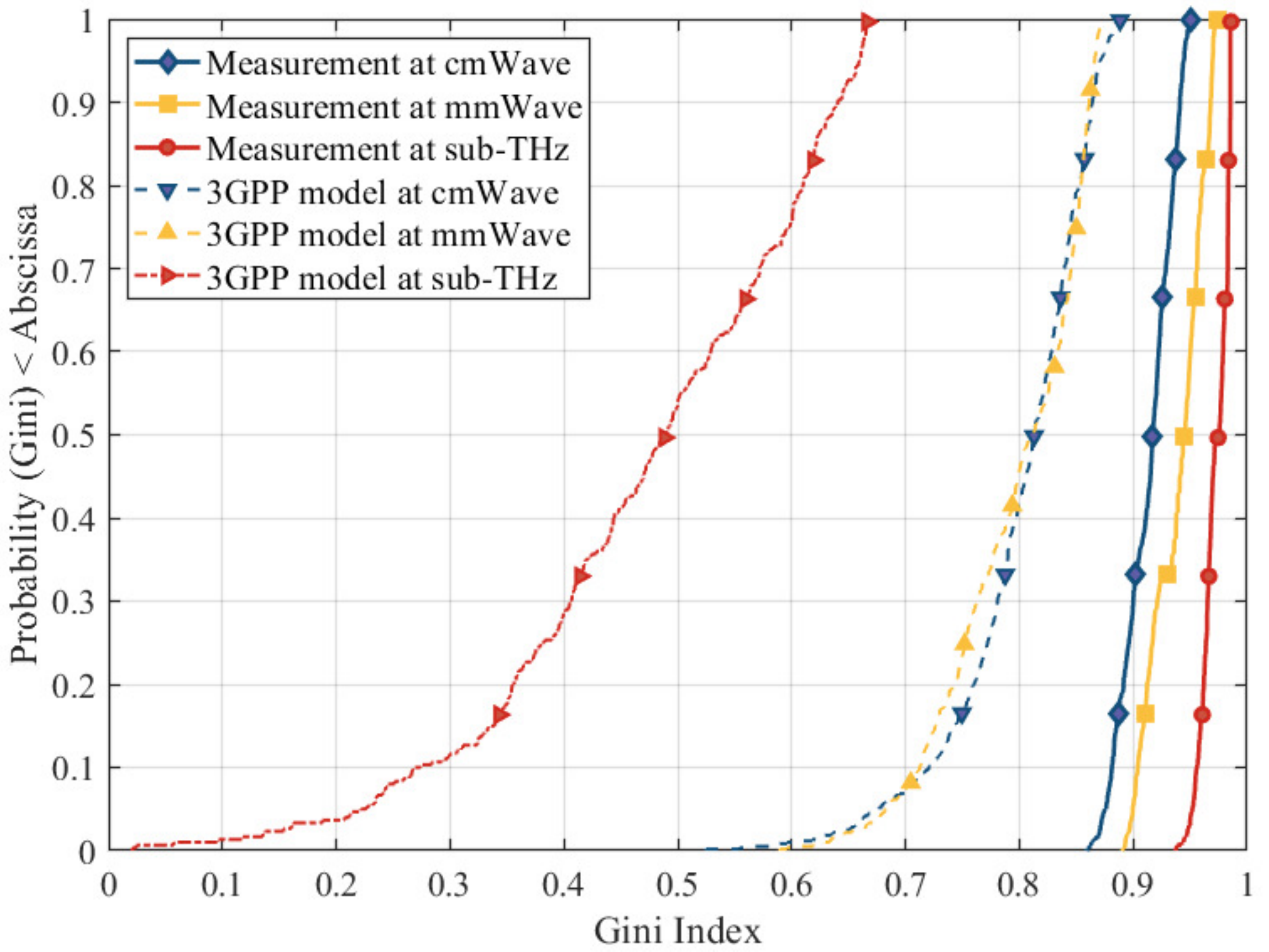}}
\caption{The CDF of Gini index with LoS path at cmWave, mmWave, and sub-THz.}
\label{Gini2}
\end{figure}

From the CDF curves of the Gini index generated by the measurement as illustrated in Fig.~\ref{Gini2}, the blue, yellow, and red lines represent the Gini index of cmWave, mmWave, and sub-THz channels, respectively. It is found that the sparsity of sub-THz channels is greater than that of the cmWave and mmWave channels, as shown by the Gini index of the sub-THz channel which is closer to 1. The mmWave channels are sparser than cmWave, though the Gini index at cmWave also results in sparsity. The Gini index of sub-THz channels is 0.96, 0.98, and 0.98 for the 20$\%$, 50$\%$, and 80$\%$ cases of the CDF, as illustrated in the measurement results from Table~\ref{Giniindex}. The Gini index at mmWave is 0.91, 0.95, and 0.96 for the 20$\%$, 50$\%$, and 80$\%$ cases of the CDF. The Gini index at cmWave is 0.89, 0.92, and 0.94 for three cases of the CDF. Since the LoS ray is the dominant ray, the sparsity of cmWave, mmWave, and sub-THz channels are significant. The significant reflection and diffraction losses in the sub-THz channels lead to the LoS path containing the vast majority of the power compared to the mmWave and cmWave channels. 

\begin{figure}[tbp]
\centerline{\includegraphics[scale=0.3]{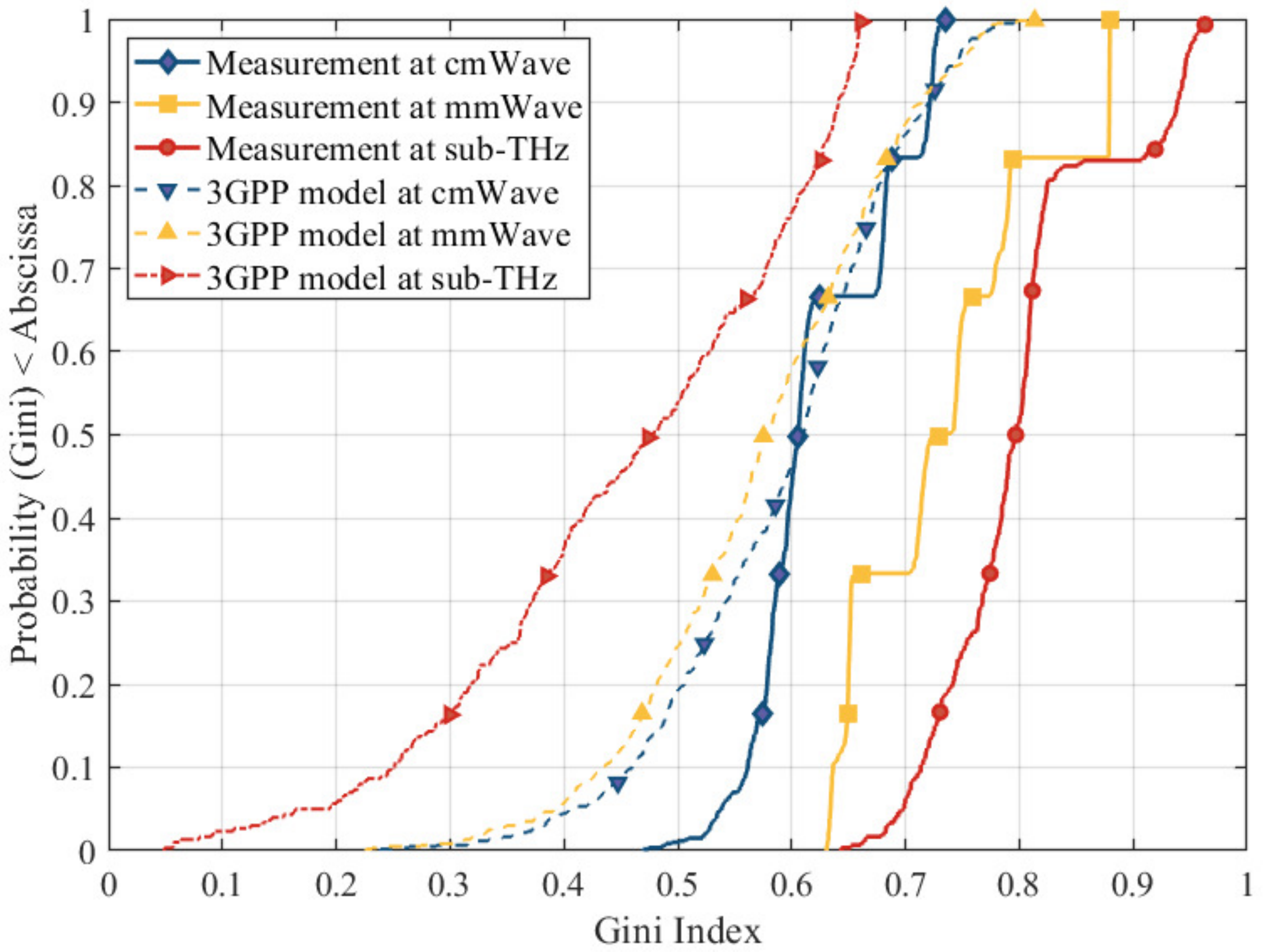}}
\caption{The CDF of Gini index without LoS path at cmWave, mmWave, and sub-THz.}
\label{withoutL}
\end{figure}

Since the LoS path contains most of the power, the Gini index of the cmWave, mmWave, and sub-THz channels are relatively similar. Therefore, we remove the LoS path to study the results of the Gini index for the remaining rays, as illustrated in Fig.~\ref{withoutL}. From the CDF curves of the Gini index without the LoS path, there are significant differences in the sparsity of different frequency bands.

From the CDF curves of the Gini index without the LoS path as illustrated in Fig.~\ref{withoutL}, the blue, yellow, and red lines represent the Gini index of cmWave, mmWave, and sub-THz channels, respectively. It is found that the channel sparsity is found to be more different for the three frequency bands. It is demonstrated that channel sparsity depends not only on the LoS path but also on the NLoS paths. In addition, the range of the CDF curves for the Gini index is much larger at cmWave, mmWave, and sub-THz. The Gini index without the LoS path is 0.74, 0.8, and 0.82 for the 20$\%$, 50$\%$, and 80$\%$ cases of the CDF in sub-THz bands, as illustrated in the Gini index without the LoS path from Table~\ref{Giniindex}. The Gini index at mmWave is 0.65, 0.73, and 0.79 for the 20$\%$, 50$\%$, and 80$\%$ cases of the CDF. The Gini index at cmWave is 0.58, 0.61, and 0.68 for three cases of the CDF. These results demonstrate that the effect of frequency on sparsity is significant. The channel sparsity in sub-THz bands is significant.

The dashed line represents the CDF curves of the Gini index derived from the simulation based on the 3GPP channel model, as illustrated by Fig.~\ref{Gini2} and Fig.~\ref{withoutL}. It is found that the Gini index derived by the 3GPP channel model is much smaller than that calculated by the measurement. It proves that the 3GPP channel model is deficient in characterizing sparsity by the Gini index. In addition, the difference in channel sparsity between cmWave and mmWave channels is slight, which demonstrates that the 3GPP channel model cannot classify the sparsity characteristics of these frequency bands. From Fig.~\ref{Gini2} and Fig.~\ref{withoutL}, it is found that the Gini index of sub-THz channels is significantly smaller than that of the cmWave and mmWave. This is because the number of clusters is too less at sub-THz, as illustrated in Table.~\ref{Largescale}. The power of the clusters is correlated with the delay. When the number of clusters is decreasing, the rays with weak power are also reduced, resulting in the reduction of the Gini index. It should be noted that the simulation results are closer to the measured results in cmWave bands. This indicates that the 3GPP channel model matches better to the real channel in cmWave bands.

From Fig.~\ref{Gini2} and Table~\ref{Giniindex}, in the 50$\%$ case of the CDF, the Gini index derived by the 3GPP channel model at sub-THz is 0.49, while the Gini index of the measured channels is almost twice the simulation results, i.e., 0.98. In the cases without the LoS path, the Gini index derived by the 3GPP channel model at sub-THz is 0.48 in the 50$\%$ case of the CDF, as illustrated by Fig.~\ref{withoutL} and Table~\ref{Giniindex}. The corresponding measurement result is 0.8, and the result differs by 0.32. It proves that the 3GPP channel model lacks the ability to characterize sparsity in the delay domain.

\begin{table*}[btp]
\caption{The Gini Index of different cases.}
\begin{center}
\begin{tabular}{|p{2cm}<{\centering}|p{1.5cm}<{\centering}|p{1cm}<{\centering}|p{1cm}<{\centering}|p{1cm}<{\centering}|p{1cm}<{\centering}|p{1cm}<{\centering}|p{1cm}<{\centering}|}
\hline
\multicolumn{2}{|c|}{\multirow{2}*{Cases}}
          & \multicolumn{3}{c|}{with LoS path}       & \multicolumn{3}{c|}{without LoS path} \\
\cline{3-8} 
\multicolumn{2}{|c|}{} & 20$\%$ & 50$\%$ & 80$\%$    & 20$\%$ & 50$\%$ & 80$\%$\\
\hline
\multirow{3}*{Measurement} 
             & cmWave  &   0.89 &   0.92 &  0.94     &0.58    &0.61    &0.68 \\
\cline{2-8} 
~            & mmWave  &   0.91 &   0.95 &  0.96     &0.65    &0.73    &0.79 \\
\cline{2-8} 
~            & sub-THz &   0.96 &   0.98 &  0.98     &0.74    &0.8    &0.82 \\
\hline
\multirow{3}*{3GPP model} 
             & cmWave  &   0.78 &   0.83 &  0.86     &0.5    &0.61     &0.68 \\
\cline{2-8} 
~            & mmWave  &   0.76 &   0.82 &  0.85     &0.48    &0.58     &0.67 \\
\cline{2-8} 
~            & sub-THz &   0.36 &   0.49 &  0.61     &0.32    &0.48     & 0.61 \\
\hline
\end{tabular}
\label{Giniindex}
\end{center}
\end{table*}

\section{Intra-Cluster Power Allocation Model}\label{IV}
Section \ref{III} demonstrated the 3GPP channel model lacks the ability to characterize sparsity in the delay domain. However, the channel sparsity has been widely used in channel studies and algorithm design. This causes limitations in the application of the 3GPP channel models in algorithm design. To enable the 3GPP channel model to characterize sparsity in high-frequency bands, this section proposes a new intra-cluster power allocation model.

\subsection{Proposed Intra-cluster K-factor}\label{IVA}

The Gini index is a parameter that describes the power distribution of the effective rays in channel studies. The deficiencies of the 3GPP channel model in characterizing sparsity are reflected by the Gini index in Section \ref{III}. Therefore, we infer that the power distribution method mentioned in the fourth step of Section \ref{IIB} based on the 3GPP channel model is not reasonable. To model the intra-cluster power allocation, the measurement results are clustered and the power distribution within the cluster is observed under the real channel. The following is an example of the sub-THz channels.



\begin{figure}[tbp]
\centering
\subfigure[]{\includegraphics[scale=0.18]{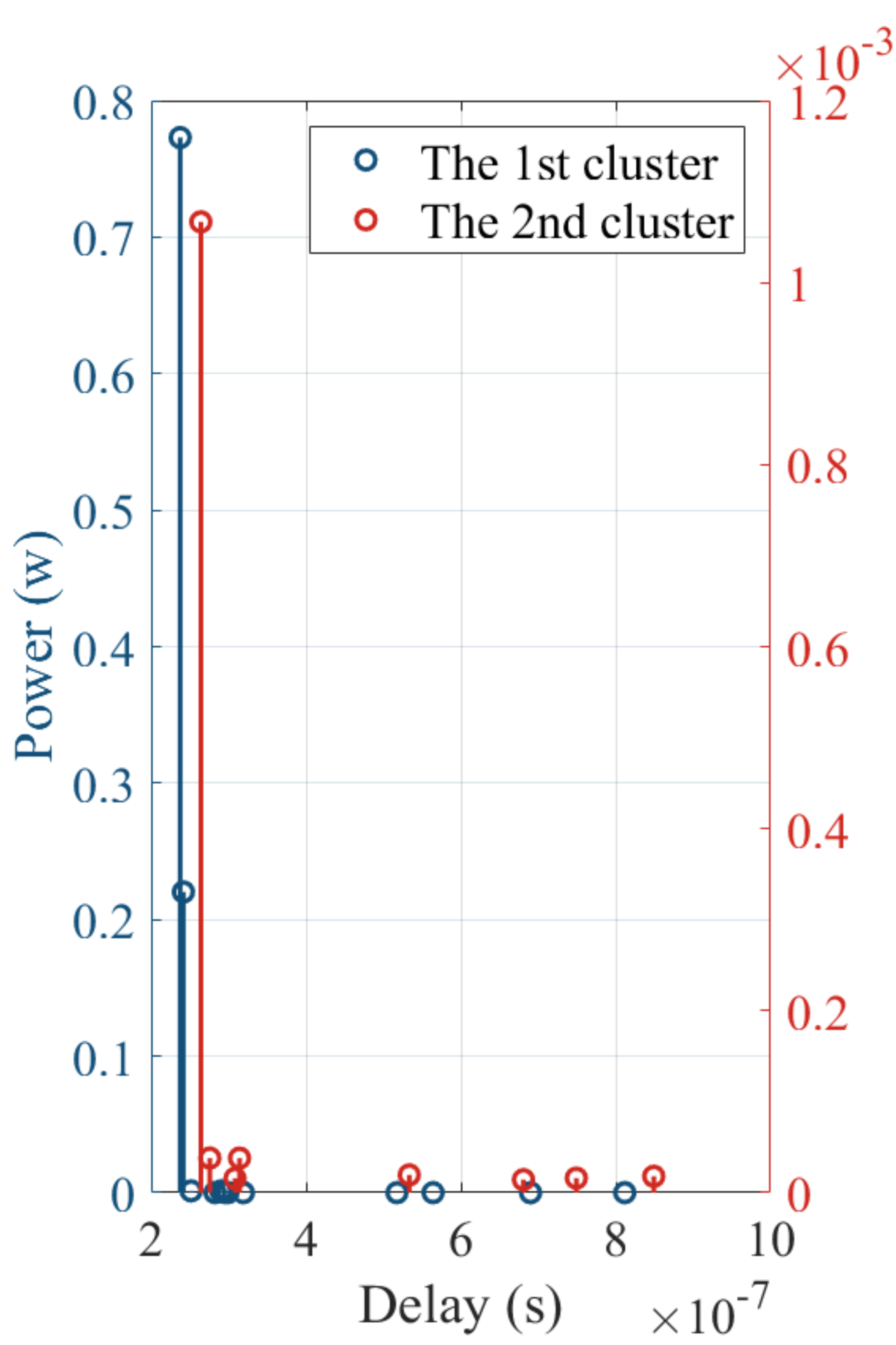}
\label{PL}}
\quad
\subfigure[]{\includegraphics[scale=0.18]{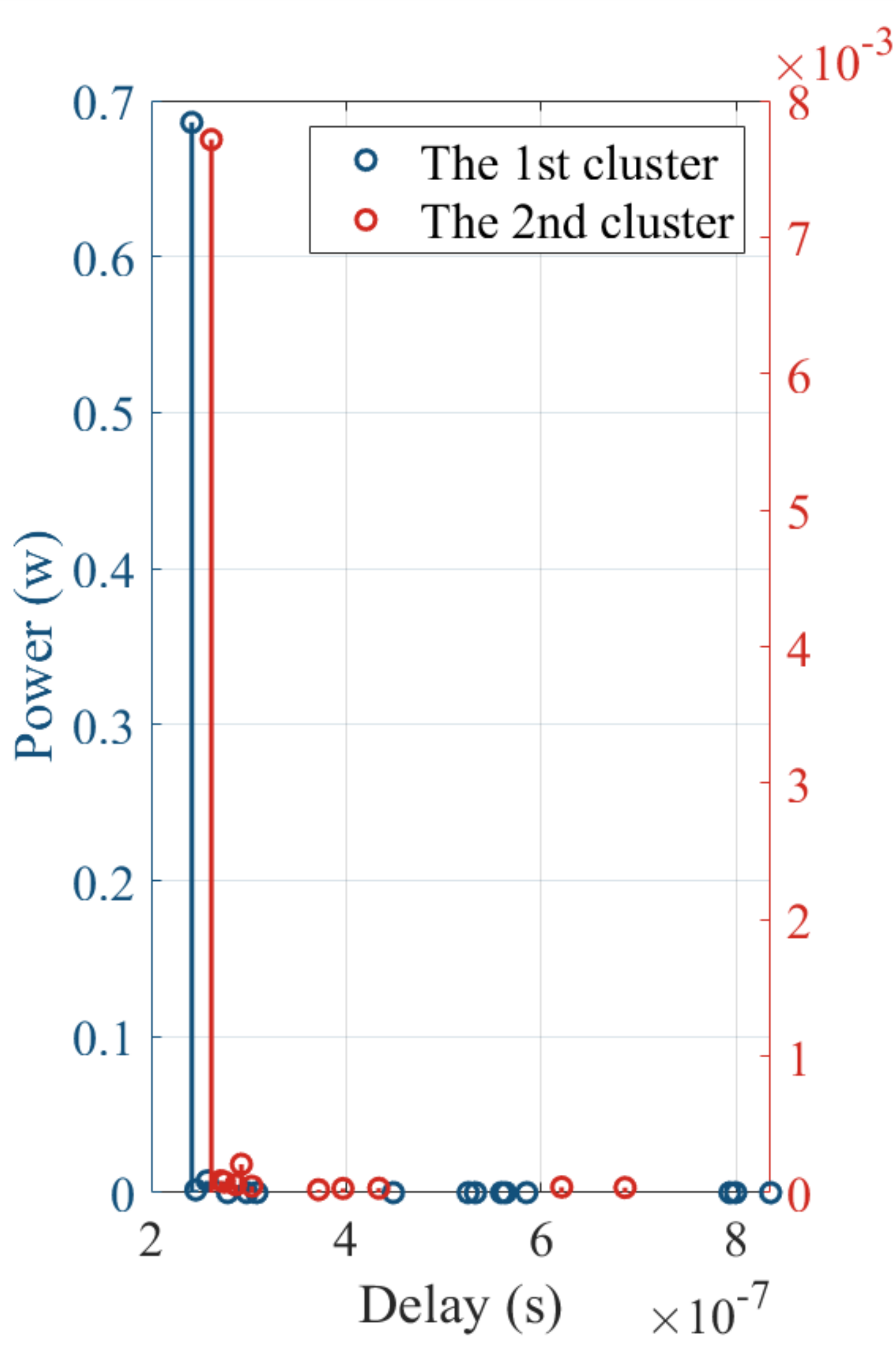}
\label{PN}}
\caption{The power distribution within the cluster of sub-THz channels. (a) The 1st position. (b) The 3rd position.}
\label{power}
\end{figure}

The effective rays of the sub-THz channels are significantly reduced, resulting in a very small number of clusters. The clustering results for most of the measured data in sub-THz bands indicate that the power distribution within the cluster shows sparse. Fig.~\ref{power} presents the intra-cluster power distribution for snapshots of the two measured positions. A complete channel is divided into two clusters in both positions, as illustrated in Fig.~\ref{PL} and Fig.~\ref{PN}. Even though the power of the two clusters differs greatly, the results indicate there is a dominant ray within each cluster. That is, a dominant ray within the cluster contains most of the power. Therefore, a new parameter, the intra-cluster K-factor (ICK), is introduced to model the power distribution within the cluster.

Based on the framework of the 3GPP channel model, the power of clusters in this method is not changed, but only the power distribution within the cluster is modified. The sparsity of the 3GPP channel model is increasing with ICK. To highlight the dominant ray in the cluster, the expression of ICK ($I$) is

\begin{equation}
I=\frac{\max\left ( \mathbf{p}_n  \right ) }{\left \| \mathbf{p}_n \right \|_1 -\max\left ( \mathbf{p}_n  \right )}, 
\label{Kc}
\end{equation}

\noindent where $\mathbf{p}_n$ represents the vector composed of the power of all rays within the $n$-th cluster.

\begin{figure}[tbp]
\centerline{\includegraphics[scale=0.2]{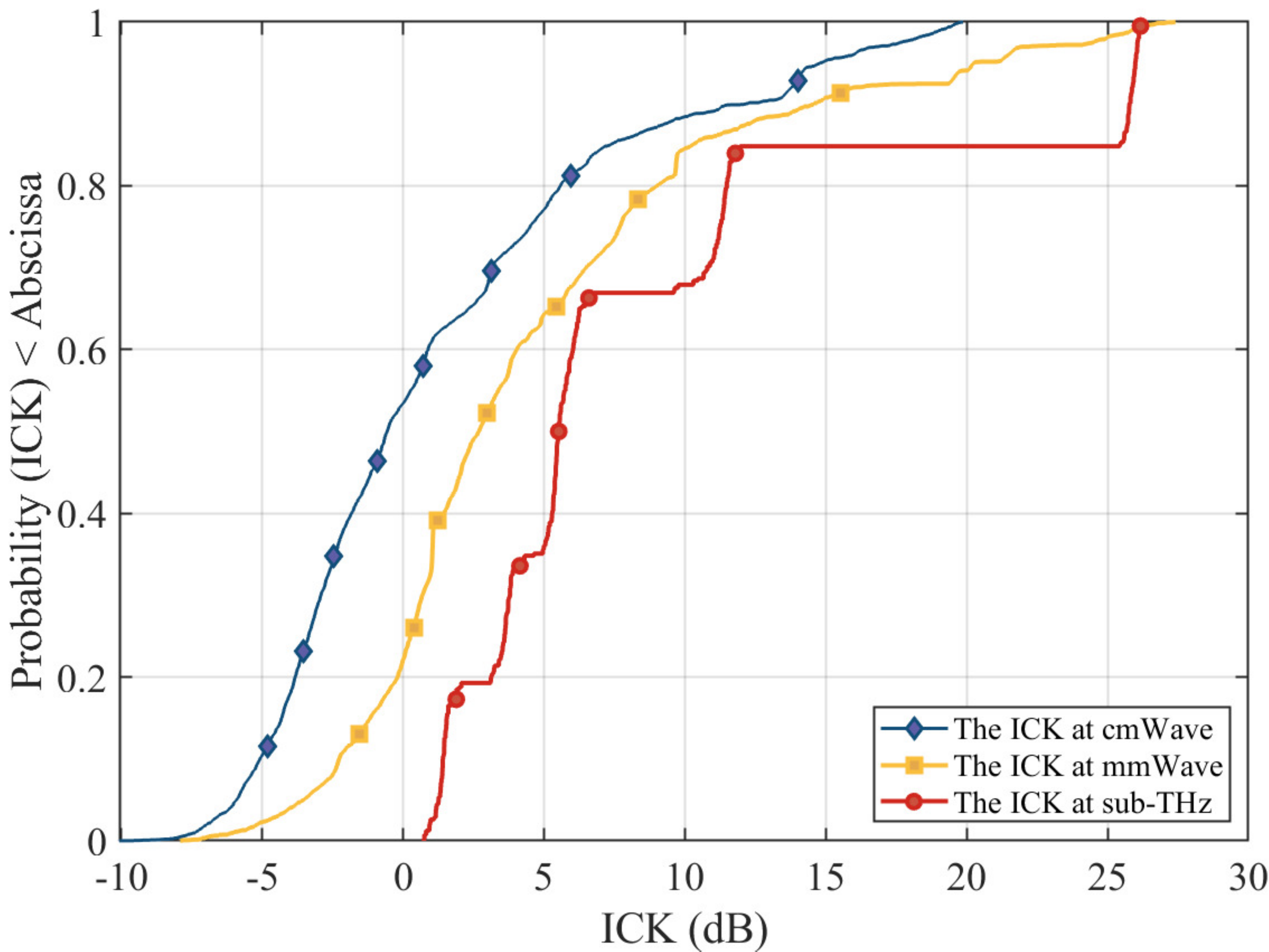}}
\caption{The CDF of ICK at cmWave, mmWave, and sub-THz.}
\label{K}
\end{figure}

The value of ICK is extracted from the measured clustering results for the statistical analysis of all positions, as illustrated in Fig.~\ref{K}. To analyze the relationship between the ICK and frequency, we use the same method to calculate the ICK for cmWave, mmWave, and sub-THz channels. From Fig.~\ref{K}, the ICK is maximum in sub-THz bands, followed by the mmWave channels. This is expected because the sparsity of the sub-THz channels is more significant. Compared to the CDF curves of the ICK in sub-THz bands, the curves at cmWave and mmWave are smoother. This is due to the the small number of clusters in sub-THz bands resulting in a small amount of data in the ICK, as illustrated in Table~\ref{Largescale}.


From Fig.~\ref{power}, a dominant ray within each cluster contains most of the power. The power of the remaining rays within each cluster is small and similar, so this paper approximates the remaining rays to be modeled as having the same power. Since the power of the cluster does not change, the sum power of all rays in the cluster remains the same. Thus, the power of the dominant ray in each cluster is

\begin{equation}
p_{n,1}=\frac{I}{I+1} \cdot P_n, 
\label{mainray}
\end{equation}

\noindent where $p_{n,1}$ represents the power of the dominant ray in the $n$-th cluster. $P_{n}$ represents the power of the $n$-th cluster. The power of the remaining rays is

\begin{equation}
p_{n,m}=\frac{1}{I+1} \cdot \frac{P_n}{M-1}, 
\label{ramainray}
\end{equation}

\noindent where $M$ represent the number of rays within a cluster, $m$ represents the index of rays within cluster, $m=2,3,\cdots ,M$. $M$ is equal to 20 in the 3GPP model. 

\subsection{Sparsity Modeling}\label{IVB}
As the frequency increases, the parameters in the 3GPP channel model differ significantly from the real channel. However, its modeling framework remains the basis for channel modeling in sub-THz communication. In channel sparsity, there is a gap between the 3GPP channel model and the real channel. This gap increases with frequency. An important reason is that the intra-cluster power mentioned in the fourth step of Section \ref{IIB} distributes the power of the cluster equally. The channel coefficients of the 3GPP channel model is expressed as \cite{b18}

\begin{equation}
H_{n} \left ( t \right ) =\sqrt{\frac{P_n}{M} } \sum_{m=1}^{M} c_{m}\exp \left ( j2\pi v_{m}t  \right ),
\label{h}
\end{equation}

\noindent where $P_n$ represents the power of the $n$-th cluster, $M$ represents the number of rays per cluster. $c_{m}$ is the complex channel coefficient. The coefficient includes the radiation pattern of the antennas, random phases and path loss. $v_{m}$ is the Doppler frequency.

The original 3GPP channel impulse responses are obtained by superimposing MPCs. The power of the cluster is first calculated using the delay of the cluster, and then the power of each ray is calculated using $P_n/M$. The intra-cluster power allocation model proposed in this paper uses ICK to change the power distribution within the cluster. Then, the channel coefficients of the $n$-th cluster can be expressed as

\begin{align}
\label{hk}
H_{n} \left ( t \right ) =\sqrt{\frac{I}{I+1}\cdot P_n} \cdot c_{1}\exp \left ( j2\pi v_{1}t  \right )\\
+\sqrt{\frac{1}{I+1}\cdot \frac{P_n}{M-1} } \sum_{m=2}^{M} c_{m}\exp \left ( j2\pi v_{m}t  \right ).
\notag
\end{align}

\section{Performance Analysis}\label{V}

In this section, a simulation experiment is designed to analyze the performance of the intra-cluster power allocation model proposed in Section \ref{IV} based on the 3GPP channel model. The design of the simulation experiments is in Section \ref{II}. In this paper, we compare with the original 3GPP channel model by changing only the power allocation model. Besides, the reliability of the intra-cluster power allocation model is verified by theoretical derivation.

\subsection{Theoretical Derivation}

In the original 3GPP channel model, the power is equally distributed within the cluster. Therefore, the Gini index calculated by the power of each ray can be expressed as

\begin{align}
\label{Giniraw}
G_1=1-\frac{2}{\left \| \mathbf{p}   \right \|_1 } \sum_{n=1}^{N}\sum_{m=1}^{M} \frac{P_n}{M }  \left ( \frac{R-M\left ( n-1 \right )-m +\frac{1}{2} }{R}  \right ),
\end{align}

\noindent where $N$ represents the number of clusters, $M$ represents the number of the rays within each cluster and $R = N\cdot M$ represents the number of all effective rays. $\mathbf{p}$ denotes the power vector of all effective rays. The elements in $\mathbf{p}$ are arranged in an ascending order, as expressed by

\begin{equation}
\mathbf{p} =\left [ \begin{matrix}
 p_{1} & p_{2} & \cdots & p_{M} & p_{M+1} & \cdots  &p_{R}
\end{matrix} \right ],
\label{p}
\end{equation}

\begin{equation}
p_{\left ( n-1 \right )M+ 1}  =p_{\left ( n-1 \right )M+ 2}= \cdots = p_{nM}=\frac{P_{n} }{M} ,
\label{vectorp}
\end{equation}

\noindent where $P_{n}$ represents the power of the $n$-th cluster, and $P_{1} <  P_{2}<  \cdots < P_{N}$.

The extended 3GPP channel model applies the power distribution model mentioned in Section \ref{IV} rather than the average distribution in the original 3GPP channel model. When the new parameter ICK is introduced, the Gini index of the channels reconstructed by the modified 3GPP channel model is expressed by

\begin{align}
\label{GiniK}
\notag
G_k=1-\frac{2}{\left \| \mathbf{d}   \right \|_1 } \left [ \sum_{n=1}^{N}\sum_{m=1}^{M-1} \frac{P_n}{\left ( I+1 \right )\left ( M-1 \right )   }  \right.\\
\phantom{=\;\;}\left.\cdot\left ( \frac{R-M\left ( n-1 \right )-m +\frac{1}2{} }{R}  \right )\right.\\
\phantom{=\;\;}\left.+\sum_{n=1}^{N} \frac{I\cdot P_n}{ I+1}  \left ( \frac{R-M\cdot n   +\frac{1}2{} }{R}  \right ) \right ] ,
\notag
\end{align}

\noindent where $\mathbf{d}$ represents the vector composed of the powers of all rays sorted in an ascending order after power distribution. In addition, $\left \| \mathbf{d}   \right \|_1 =\left \| \mathbf{p}   \right \|_1 $, because the cluster power is the same as before. The number of rays $R$ remains the same because the number of clusters and rays is the same.

After the power distribution with ICK, two situations may occur:

\begin{itemize}
\item The sorting of all elements in $\mathbf{d}$ is the same as $\mathbf{p}$. That is, the power of the dominant ray in $P_{i}$ is still less than the power of the remaining rays in $P_{i+1}$ after the power distribution.
\item The sorting of all elements in $\mathbf{d}$ is different from $\mathbf{p}$. That is, the power of the dominant ray in $P_{i}$ is greater than the power of the remaining rays in $P_{i+1}$ after the power distribution.
\end{itemize}

Since the power of the clusters is not changed, the power distribution within the cluster can be regarded as giving the power of the remaining rays to the dominant ray. We define the power given out as $\alpha ,\cdots ,\beta $. When the order of the ray power is constant, the elements of the power vector $\mathbf{d}$ are expressed as 

\begin{align}
\label{vectord}
\notag
d_{i} & = p_{i} -\alpha ,i=1,2,\cdots,M-1\\
\notag
d_{M} & = p_{M} +\left ( M-1 \right ) \alpha\\
d_{j} & = p_{j} -\beta  ,j=M+1,\cdots,R-1\\
\notag
d_{R} & = p_{R} +\left ( M-1 \right )\beta 
\end{align}

Eq.~\eqref{GiniK} is rewritten by bringing Eq.~\eqref{vectord} into Eq.~\eqref{GiniK}. Then, the result of Eq.~\eqref{GiniK} minus Eq.~\eqref{Giniraw} can be regarded as the superposition of $N$ similar parts. Eq.~\eqref{part1} is the expression for $i=1,2,\cdots ,M$. The rest of the expressions only change the power given out $\alpha$. Each part is proved to be a positive number, expressed as

\begin{align}
\label{part1}
\notag
&2\sum_{i = 1}^{M-1} \frac{\alpha }{\left \| \mathbf{p}  \right \|_{1}  } \left ( \frac{R-i+\frac{1}{2} }{R}  \right ) -\frac{\left ( M-1 \right )\alpha  }{\left \| \mathbf{p}  \right \|_{1}}\left ( \frac{R-M+\frac{1}{2} }{R}  \right ) \\ 
=&\frac{\alpha }{\left \| \mathbf{p}  \right \|_{1}  }\cdot \frac{M^{2}-M }{R} > 0.
\end{align}

\begin{figure}[tbp]
\centerline{\includegraphics[scale=1]{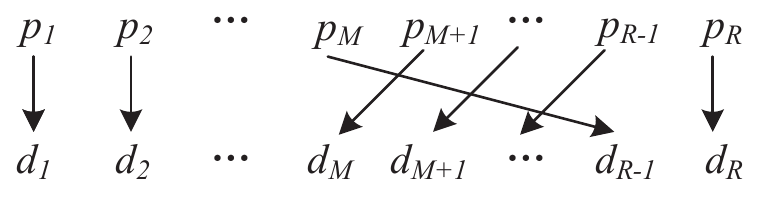}}
\caption{The order of the elements in the power vector $\mathbf{d}$.}
\label{order}
\end{figure}

The order of the ray power in $\mathbf{d}$ is changed in the second situation. Taking the case of two clusters as an example, the power of the rays in the vector $\mathbf{d}$ after power allocation is the same as in the first situation. But the order of the elements has changed, as illustrated in Fig.~\ref{order}. In the vector $\mathbf{p}$, the power of the second cluster is greater than that of the first cluster. The power of the rays in each cluster is the same, i.e., $p_{1} = \cdots =p_{M} < p_{M+1}= \cdots =p_{R}$. After power distribution, when the power of the dominant ray in the first cluster is greater than the remaining rays in the second cluster, the subscript of the dominant ray in the first cluster becomes larger, i.e., $R-1$. The elements of the power vector $\mathbf{d}$ are expressed as 

\begin{align}
\label{vectord2}
\notag
d_{i} & = p_{i} -\alpha ,i=1,2,\cdots,M-1\\
\notag
d_{j} & = p_{j+1} -\beta,j=M,M+1,\cdots,R-2\\
d_{R-1} & = p_{M} +\left ( M-1 \right )\alpha\\
\notag
d_{R} & = p_{R} +\left ( M-1 \right )\beta 
\end{align}

Eq.~\eqref{GiniK} is rewritten by bringing Eq.~\eqref{vectord2} into Eq.~\eqref{GiniK}. Then, the result of Eq.~\eqref{GiniK} minus Eq.~\eqref{Giniraw} can be expressed as

\begin{align}
\label{difference}
\notag
G_{k}-G_{1} = &2 \sum_{i = 1}^{M-1} \frac{\alpha }{\left \| \mathbf{p } \right \|_{1} }\left ( \frac{R-i+\frac{1}{2} }{R}  \right )\\
\notag
&+ 2 \sum_{i = M}^{N-2} \frac{p_{i}-p_{i+1}+ \beta  }{\left \| \mathbf{p } \right \|_{1} }\left ( \frac{R-i+\frac{1}{2} }{R}  \right )\\
&+2\cdot \frac{p_{R-1}- p_{M}-\left ( M-1 \right )\alpha  }{\left \| \mathbf{p}  \right \|_{1}  }\cdot  \frac{3}{2R} \\
&-\frac{\left ( M-1 \right ) \beta }{\left \| \mathbf{p}  \right \|_{1} } \cdot  \frac{1}{2R}.
\notag
\end{align}

Eq.~\eqref{difference} is easily proven to be a positive number. Therefore, $G_k$ is greater in both situations, i.e., the sparsity is enhanced in the extended 3GPP model with the proposed intra-cluster power allocation model. The reference values of ICK are given in Table~\ref{Kcmea} based on the measured data in cmWave, mmWave, and sub-THz bands.


\begin{table}[htbp]
\setlength{\tabcolsep}{7mm}
\caption{The reference value of ICK}
\begin{center}
\begin{tabular}{|p{1.5cm}<{\centering}|p{3.5cm}<{\centering}|}
\hline
Frequency& Reference value of ICK (dB)\\
\hline
cmWave   & 4.93\\
\hline
mmWave   & 9.86\\
\hline
sub-THz  & 17.99\\
\hline
\end{tabular}
\label{Kcmea}
\end{center}
\end{table}



\subsection{Simulation Results Analysis}

\begin{figure}[tbp]
\centering
\subfigure[]{\includegraphics[scale=0.3]{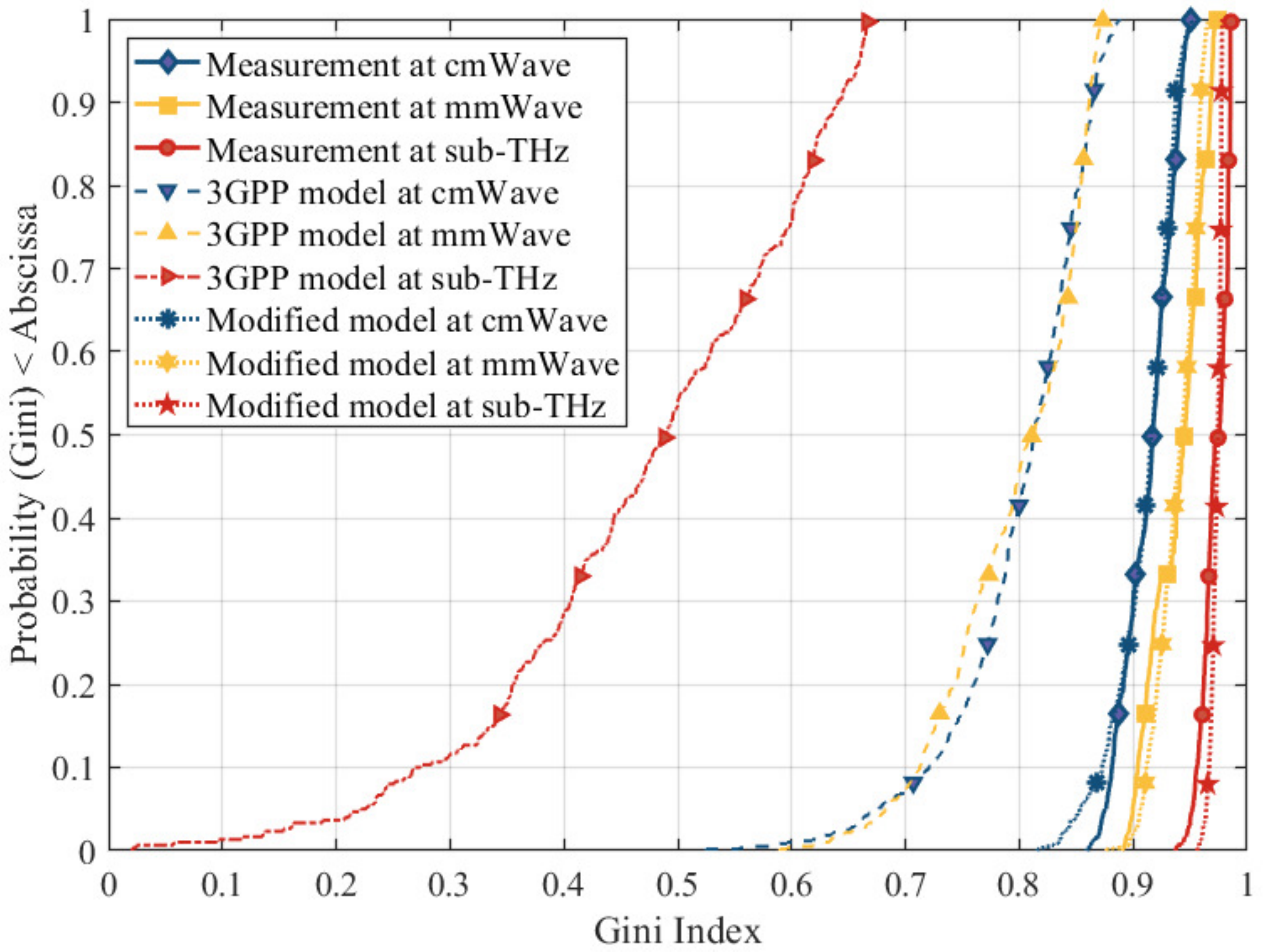}
\label{CL}}
\subfigure[]{\includegraphics[scale=0.3]{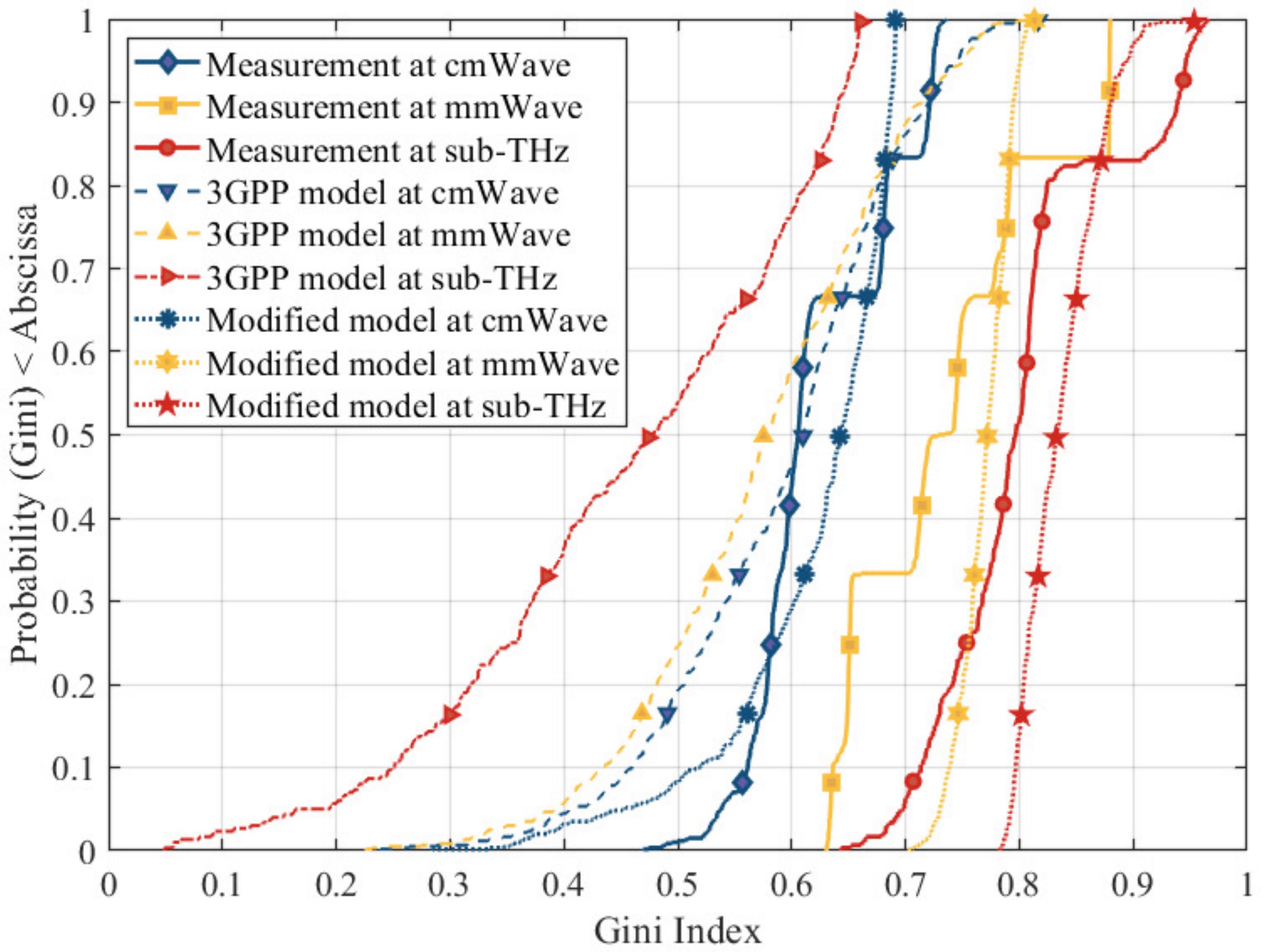}
\label{CNL}}
\caption{The CDF of Gini index generated by measured channels, by the 3GPP channel model, and by the 3GPP model with proposed intra-cluster power allocation model (modified model). (a) With LoS path. (b) Without LoS path.}
\label{sparsitym}
\end{figure}

The simulation experiment reconstructs the Gini index of the channels based on the 3GPP channel model framework. This paper compares the simulation results between the original simulation condition and after introducing the intra-cluster power allocation model. The reliability of the intra-cluster power allocation model is analyzed by comparing them with the measurement results, as illustrated in Fig.~\ref{sparsitym}.


Comparing the Gini index generated by measured channels, by the 3GPP channel model, and by the modified 3GPP model with the proposed intra-cluster power allocation model, it is found that the sparsity of the modified model is more significant than that of the 3GPP model in cmWave, mmWave, and sub-THz bands, and matches the measured results better, as illustrated in Fig.~\ref{sparsitym}. This proves that the proposed intra-cluster power allocation model performs well both with and without the LoS path. The LoS path is an important factor affecting channel sparsity, so we discuss the performance of the proposed intra-cluster power allocation model with and without the LoS path, respectively. 

From Table~\ref{Giniindex} and Table~\ref{THGini}, it is found that the measured Gini index with the LoS path in the 50$\%$ case of the CDF is 0.98 in sub-THz bands, while the Gini index of the 3GPP model is 0.49, which differs from the measured results by 0.49. In the 50$\%$ case of the CDF, the Gini index of the modified model with the LoS path is 0.97 in sub-THz bands, which differs from the measured results by 0.01. In mmWave bands, the Gini index of the modified model with the LoS path is 0.94 in the 50$\%$ case, which differs from the measured results by 0.01. While the Gini index of the original 3GPP model with LoS path is 0.82, which differs from the measured results by 0.13. In cmWave bands, the Gini index generated by the modified 3GPP channel model with the intra-cluster power allocation model is 0.92, while the Gini index is 0.83 for the 3GPP channel model. These results prove that the original 3GPP model lacks the ability to characterize sparsity in the delay domain and the modified 3GPP model addresses the deficiency. 

In the absence of the LoS path, the modified model with the intra-cluster power allocation model also addresses the deficiencies of the 3GPP channel model, as illustrated in Fig.~\ref{CNL}. In sub-THz bands, the Gini index of the modified model without the LoS path is 0.83 in the 50$\%$ case of the CDF, which differs from the measured results of less than 0.03. While the Gini index of the 3GPP model is 0.48, which differs from the measured results by 0.32. This result demonstrates that the modified model with the intra-cluster power allocation model enables the 3GPP channel model to characterize sparsity in the delay domain at cmWave, mmWave, and sub-THz. 

From the CDF curves of the 3GPP model results illustrated in Fig.~\ref{sparsitym}, it is found that the difference in channel sparsity between cmWave and mmWave channels is slight. This may be due to the similar number of clusters. It is demonstrated that the 3GPP channel model cannot distinguish sparsity by frequency. Nonetheless, the measurements prove that sparsity is strongly related to frequency. This is an important deficiency of the 3GPP model in describing sparsity. The CDF curves of the Gini index derived from the modified model show the differences between frequencies even in the case of a similar number of clusters.

\begin{table}[htbp]
\caption{The Gini index derived by the modified 3GPP model at cmWave, mmWave, and sub-THz}
\begin{center}
\begin{tabular}{|p{0.8cm}<{\centering}|p{1.3cm}<{\centering}|p{1cm}<{\centering}|p{1cm}<{\centering}|p{1cm}<{\centering}|}
\hline
\multicolumn{2}{|c|}{Cases}   & 20$\%$ & 50$\%$  &  80$\%$\\
\hline
\multirow{3}{*}{\makecell[c]{with \\ LoS \\ path }}
&          cmWave             & 0.89   &  0.92   &  0.93\\
\cline{2-5}
&          mmWave             & 0.92   &  0.94   &  0.96\\
\cline{2-5}
&          sub-THz            & 0.96   &  0.97   &  0.98\\
\hline
\multirow{3}{*}{\makecell[c]{without \\ LoS \\ path }}
&          cmWave             & 0.57   &  0.64   &  0.68\\
\cline{2-5}
&          mmWave             & 0.75   &  0.77   &  0.79\\
\cline{2-5}
&          sub-THz            & 0.8   &  0.83   &  0.87\\
\hline
\end{tabular}
\label{THGini}
\end{center}
\end{table}

\section{Conclusion}\label{VI}

In this paper, the channel sparsity is analyzed and modeled in cmWave, mmWave, and sub-THz bands by multi-band measurements. The channel measurements at 6, 26, and 132 GHz were performed in the same indoor office scenario to analyze the channel sparsity by Gini index. It is proved that the sparsity of the sub-THz channels is more significant than that of cmWave and mmWave by comparing the Gini index. In the 50$\%$ case of the CDF, the Gini index obtained by measurement with the LoS path is 0.98 in sub-THz bands. The Gini index is 0.95 in mmWave and 0.92 in cmWave bands. The underlying framework of the 3GPP channel model is still applicable to high-frequency communications. However, the parameters in the model need to be corrected with a higher carrier frequency. Therefore, this paper used large-scale parameters which are obtained by measurement in the 3GPP model to obtain the Gini index of the channels at the three frequency bands. The comparison results demonstrate the weak ability of the 3GPP channel model to characterize sparsity in the delay domain. The measured Gini index in the 50$\%$ case of the CDF is 0.98 in sub-THz bands, while the simulation results based on the original 3GPP channel model is only 0.49. 

A new intra-cluster power allocation model is proposed based on the measurement. The modified 3GPP channel model with the proposed intra-cluster power allocation model has the ability to characterize channel sparsity in the delay domain, which is verfied by theoretical derivation and simulation experiments. The reference values of ICK from the derived intra-cluster power allocation model are given in Table~\ref{Kcmea}. The reference value is 4.93, 9.86, and 17.99 dB in cmWave, mmWave, and sub-THz bands, respectively. In the 50$\%$ case of the CDF, the Gini index obtained by the modified model with new intra-cluster power allocation is 0.97 in sub-THz bands, which differs from the measured results by only 0.01. The intra-cluster power allocation model also performs well in the cmWave and mmWave bands. These results demonstrate that the proposed models are suitable for 3GPP-type channels. In the future, we will continue to study the channel sparsity of multiple input multiple output systems and model sparsity using multiple metrics.

\section*{Acknowledgments}

This research is supported in part by National Natural Science Foundation of China (61925102 \& 92167202 \& 62201086 \& 62101069), National Key R\&D Pro-gram of China (2020YFB1805002), and Beijing University of Posts and Tele-communications-China Mobile Research Institute Joint Innovation Center.


\bibliographystyle{IEEEtran}
\bibliography{ref}
\end{document}